\input harvmac.tex
\let\d\partial
\def\text#1{\quad\hbox{#1}\quad}
\def\frac#1#2{{#1 \over #2}}
\def\com#1#2{\bigl[\,  #1\,  , \,  #2  \, \bigr] }
\def\comg#1#2{\Bigl[\,   #1\,  , \,   #2  \, \Bigr] }

\def\ket#1{\ | \,  #1  \!  >\ }
\def\bra#1{\ < #1 \, |\ }
\def\bracket#1#2{\!< #1\, |\, #2\! >\, }
\def\b#1{{\bf #1}}

\def\kdots{\!\! ... \,}
\def\sgn#1{\hbox{sgn}(#1)}
\def\detq#1{\hbox{Det}_{\hbox{\scriptsize q}}\bigl[#1\bigr]}
\def\sym{\ket{{\hbox{sym}}\hbox{\scriptsize (}\b{r}\hbox{\scriptsize )}}}

\def\somprim{\mathop{\sum \!{}^{{}^{\scriptstyle \prime}}}}
\def\sump#1#2{\somprim_{\! #1}^{\! \hbox{\raise -0.65pt\hbox{$\scriptstyle
#2$}}}}
\def\st{\scriptstyle}
\def\sst{\scriptscriptstyle}

\newcount\eqnum
\eqnum=0
\def\eq{\eqno(\secsym\the\meqno)\global\advance\meqno by1}
\def\eqlabel#1{{\xdef#1{(\secsym\the\meqno)}}\eq }
\def\eqnumlabel#1{\xdef#1{(\secsym\the\meqno)}%
           &(\secsym\the\meqno)\global\advance\meqno by1}

\newwrite\refs
\def\startreferences{
 \immediate\openout\refs=references
 \immediate\write\refs{\baselineskip=14pt \parindent=16pt \parskip=2pt}
}
\startreferences

\refno=0
\def\aref#1{\global\advance\refno by1
 \immediate\write\refs{\noexpand\item{\the\refno.}#1\hfil\par}}
\def\ref#1{\aref{#1}\the\refno}
\def\refname#1{\xdef#1{\the\refno}}

\newcount\exno
\exno=0
\def\Ex{\global\advance\exno by1{\noindent\sl Example \the\exno:

\nobreak\par\nobreak}}

\parskip=6pt

\overfullrule=0mm

\font\scriptsize=cmr7


\Title{\vbox{\baselineskip12pt\hbox{LAVAL-PHY-{00-21}}}}
  {\vbox{\centerline{Conserved charges of non-yangian type}
 \smallskip {\centerline{for the Frahm-Polychronakos spin chain}}}}

\smallskip
\centerline{ P. Mathieu and Y. Xudous \foot{Work
supported by NSERC (Canada) and FCAR (Qu\'ebec) } }

\smallskip\centerline{ \it D\'epartement de
Physique,} \smallskip\centerline{Universit\'e Laval,}
\smallskip\centerline{ Qu\'ebec, Canada G1K 7P4}
\smallskip\centerline{pmathieu@phy.ulaval.ca, yxudous@phy.ulaval.ca}
\vskip .2in
\bigskip
\bigskip
\centerline{\bf Abstract}
\bigskip
\noindent
Through an $\hbar$-expansion of the confined Calogero model with spin
exchange interactions, we
extract a generating function for the involutive conserved charges of the
Frahm-Polychronakos spin chain.
The resulting conservation laws possess the spin chain yangian symmetry,
although they are not expressible in terms of these yangians.
\leftskip=0cm \rightskip=0cm
\Date{08/00 (revised: 02/01)\ }


\newsec{Introduction}

Integrable spin chains with long-range
interactions have remarkable properties, not the least being that they
furnish a sort of discretization of particular conformal field theories
with Lie group
symmetry [\ref{F.D.M. Haldane, Z.N.C. Ha, J.C. Talstra, D. Bernard and V.
Pasquier,
Phys. Rev. Lett. \b{69}  (1992) 2021}\refname\pasquier, \ref{D. Bernard, V.
Pasquier and
D. Serban, Nucl. Phys. {\bf B 428} (1994) 612.}\refname\serban, \ref{K.
Schoutens, Phys.
Lett. {\bf B 331} (1994) 335; P. Bouwknegt. A.A. Ludwig and K. Schoutens, Phys. Lett.
{\bf B 359} (1995) 304.}\refname\schou]. The archetypal model  is the
Haldane-Shastry
model [\ref{F.D.M. Haldane, Phys. Rev. Lett. {\bf 60} (1988) 635; B.S.
Shastry, Phys.
Rev. Lett. {\bf 60}  (1988) 639.}]  in which $N$  $su(n)$ spins  placed
equidistantly on a circle are coupled by a spin-exchange interaction
proportional to the
inverse square of their chord distance:
$$
         H^{\hbox{\scriptsize (HS)}}=\frac{1}{2}\sump{i, j=1}{N}\frac{z_iz_j}
{z_{ij}z_{ji}}{P}_{ij}\,  .
\eqlabel\hs$$
Here,  $z_j \equiv \exp{\left(i \frac{2 \pi j  }{N}\right)}$,
 $z_{ij} \equiv z_i-z_j$ and $P_{ij}$ is the operator which
exchanges the  $i^{\hbox{\scriptsize  $\, \!\!$ th}}$ and
$j^{\hbox{\scriptsize $\,  \!\!$  th}}$ spins. The primed sum indicates
that the summation variables are restricted to differing values.

 The Haldane-Shastry model possesses a yangian symmetry algebra
which can be taken as  a manifestation of its integrability [\pasquier].
The conserved charges directly associated to this symmetry are not scalar
(they transform in the fundamental representation of $su(n)$) and do not
commute among
themselves (they generate the yangian algebra, which is non-Abelian).
However, from
these charges,  one can build a set of $N$ scalar  commuting operators
which turns out
to be directly related to those obtained in [\ref{M. Fowler and J. A.
Minahan, Phys.
Rev. Lett. \b{70}  (1993) 2325}\refname\fowler].   However, this set does
not explicitly
contain  the Hamiltonian,  contrary to the natural expectation. Moreover,  two
additional conservation laws were known from brute force calculations
[\pasquier,
\ref{V.I. Inozemtsev, J. Stat. Phys. {\bf 59} (1990) 1143.}] but did not
appear in this
sequence. One expects that, together with $ H^{\hbox{\scriptsize (HS)}}$,
these represent
the first few of a new sequence of a conserved charges.  It is natural to
try to fit
this other sequence in a general scheme based on the  fundamental object at
the root of
integrability : the monodromy matrix. For the Haldane-Shastry model, this
has been
accomplished by  Haldane and Talstra  [\ref{J.C. Talstra and F.D.M.
Haldane, {\it
Integrals of motion of the Haldane-Shastry model}, cond-mat/9411065
(1994).}\refname\talhal]. They showed that the `new'  conservation  laws
can be obtained
by taking a rather subtle limit  of the more general dynamical spin model.

For the well-known XXX model, which has short-range interactions,
there are also two sets of conservation laws: there is a yangian symmetry
[\ref{D.
Bernard, Int. J. Mod. Phys. {\bf B7} (1993) 3517}\refname\bernard], out of
which scalar
conservation laws can be constructed and, in addition, there is a sequence of
conservation laws that includes the Hamiltonian [\ref{E.K. Sklyanin, L.A.
Takhtajan and
L.D. Faddeev, Theor. Math. Phys. \b{40} (1980) 688; L.A. Takhtajan and L.D.
Faddeev,
Russ. Math. Surv. \b{34} (1979) 11; V.E. Korepin, N.M. Bogoliubov and A.G.
Izergin, {\it
Quantum inverse scattering method and correlation functions}, Cambridge
Univ. Press,
London (1993).}]. These two types of conservation laws are easily
distinguished in
models with short-range interactions: the first set is non-local (i.e., the
conserved
charges involve interactions of all the spins and they become truly
non-local in the
continuum limit), while the set containing the Hamiltonian is local (i.e.,
the $n$-th
member of this sequence has a leading term describing the interaction of
$n$ adjacent
sites).

For spin chains with long-range interactions, the distinction between
locality and non-locality is rather artificial, both sets of charges being
manifestly non-local.
 The difference between these two sets lies in the fact that the
Hamiltonian set found by Haldane
and Talstra commutes with the symmetry algebra while the  yangian set does
not. Since both sets
commute and can therefore be simultaneously diagonalized, this means that
the eigenvalues
of the Hamiltonian set are degenerate and characterize a given multiplet
while those of the yangian
set can be used to label the differing states inside the multiplet.

Let us point out, {\it en passant}, another major difference between
integrable long- and short-range interacting chains, apart from the
relativity of the
locality concept. For short-range interacting chains, there exists a boost
operator that
allows for a recursive construction of the local conservation laws.  Its
origin can
actually be traced back to  the transfer matrix formalism and the locality
of the
interaction [\ref{M.G. Tetelman, Sov. Phys. JETP {\bf 55} (1981) 306}].  No such
operator is known for long-range  interacting chains.

The argument of [\talhal] relies on a limiting formulation of
the Haldane-Shastry spin chain.  The model can be viewed as a special
reduction of a
general Sutherland model (a dynamical model with $\sin^{-2} r$ interaction)
with spin
degrees of freedom.

The introduction of the spin degrees of freedom in a Calogero-Moser-Sutherland
model is rather direct [\ref{A.P. Polychronakos,
 Phys. Rev. Lett. \b{69} (1992) 703.}\refname\polyexch] (see also [\ref{B.
Sutherland and
B.S. Shastry, Phys. Rev. Lett. \b{71} (1993) 2329.}]) . If in the classical
version of the
model, the  potential takes the form
$\sum g f(r_i,r_j)$ (up to a possible harmonic part), where $g$ is a
coupling constant,
the quantum version reads $\sum g(g+1) f(r_i,r_j)$. The integrability turns
out to be preserved
if the term $g(g+1)$ is replaced by $g(g+K_{ij})$ where $K_{ij}$
interchanges the positions
$i$ and $j$.  The spin degrees of freedom can be introduced directly by
imposing the $K_{ij}$
to be a spin-exchange  instead of a position-exchange operator.  Another
approach, albeit less
direct, amounts to retain the position meaning of $K_{ij}$ but consider
states that are symmetric
under the interchange of both the position and the spin variables.  The
resulting effect is identical.

The transition from a dynamical model with spin degrees of freedom to the
spin chain has been phrased  in general terms by Polychronakos in [\ref{A.P.
Polychronakos,  Phys. Rev. Lett. \b{70} (1993) 15}\refname\polylat]. The
idea is simply
that from a dynamical model with spin degrees of freedom, we can somehow
freeze the
latter to generate a spin chain.  However, this freezing entails a compatibility
condition that follows from the original equations of motion:  the position
variables
must correspond to the zeroes of the potential.   For the $\sin^{-2} r$
interaction
potential, this fixes the positions of the chain sites to the roots of
unity.  Note, on
the other hand, that if the potential contains an harmonic piece, this part
does not
contribute to the spin interaction potential but it enters in the
definition of the
minima (in fact, whenever it is present, it ensures the existence of these
minima).

In this letter, we study the Hamiltonian  conservation laws of the
Frahm-Polychronakos spin chain [\polylat,
\ref{H. Frahm, J. Phys. \b{A26} (1993) L473.}\refname\frahm].  It
originates from a Calogero model with inverse square interaction and an
harmonic confining potential, augmented with spin degrees of freedom. The
potential minima fix the sites of the chain to correspond to the zeroes of
the Hermite polynomial $H_N(x)$, to be denoted $x_i$.  The Hamiltonian
takes the form
$$
     H^{\hbox{\scriptsize (FP)}}=\frac{1}{2}\sump{i,
j=1}{N}\frac{1}{x_{ij}x_{ji}}{P}_{ij}
\eqlabel\fp$$
and we will consider the general case of $su(n)$ spins, each
of the $N$ spins belonging to the fundamental representation.  This model
has already been
shown  to be integrable and to possess a yangian set of commuting operators
[\polylat].
In the following,  we will show that  Haldane-Talstra's argument,
formulated here in
a somewhat different way, can also be successfully applied to this model,
effectively
generating the set of Hamiltonian conservation laws.

\newsec{Integrability and Conservation laws}
\subsec{The yangian algebra $Y[su(n)]$}

Let us first briefly review the Yangian algebra $Y[su(n)]$
(see for instance [\bernard, \ref{F.D.M. Haldane, {\it Physics of the ideal semion gas:
spinons and quantum symmetries of the integrable Haldane-Shastry
model\/},cond-mat/9401001 (1994)}\refname\semion]) focusing on its relation
to the
construction of commuting invariants. For all  known integrable spin chains
(except, in
fact, for a single and somewhat pathological  example [\ref{M. Grabowski
and P. Mathieu,
J. Math. Phys. {\bf 36} (1995) 5340.}]), the integrability property can be
traced back
to the existence of a monodromy matrix, an $n \times n$  matrix of operator
entries
which depends on a spectral parameter $u$ and which satisfies the RTT relation:
$$
  \b{R} (u-v )\b{T^{ (1)}} (u)\b{T^{(2)}} (v)= \b{T^{(2)}}
(v)\b{T^{ (1)}}(u) \b{R} (u-v)\, .
\eq$$
Here, the  superscripts refer to two auxiliary subspaces in
which the matrices act non-trivially, e.g.
$$\b{T^{(1)}}(u) \equiv \b{T}(u) \otimes \b{1_{n \times n}}\eq$$ and
$\b{R}$, called the R-matrix, is an $n^2 \times n^2$
c-number matrix which  must satisfy the quantum Yang-Baxter equation:
$$
          \b{R^{(12)}}(u) \b{R^{(13)}} (u+v)\b{R^{(23)}}(v)=
\b{R^{(23)}}(v) \b{R^{(13)}}(u+v)\b{R^{(12)}}(u)\, .
\eq$$
The RTT relation ensures that the transfer matrix $\b{t}(u)$, which is
defined as the trace of the monodromy matrix $\b{t}(u)\equiv
\sum_{a=1}^nT^{aa}(u)$, satisfies
$$
[\b{t}(u),\b{t}(v)]=0
\eq$$
so that its expansion in power series in $u^{-1}$ generates commuting
conserved quantities (see below). The Yang-Baxter relation is simply a
compatibility relation for the RTT relation.

A simple solution to the Yang-Baxter equation is given by Yang's rational
solution
$$
    \b{R^{(ij)}} (u)=u+\lambda \b{P^{(ij)}}\, ,
\eq$$
 where $\lambda$ is an unspecified deformation parameter and
$\b{P^{(ij)}}$ exchanges the auxiliary subspaces $i$ and $j$
$$
\b{P^{(12)}} \b{A^{(1)}}\b{B^{(2)}}=\b{B^{(1)}}\b{A^{(2)}}\b{P^{(12)}}\, .
\eq$$
With  this  choice of R-matrix and with  the monodromy matrix expanded in a
Laurent series as (denoting the $ab$ matrix entry of $\b{T}(u)$ as
$T^{ab}$)
$$
     T^{ab}(u)=\delta^{ab}+\lambda\sum_{m=0}^{\infty}u^{-(m+1)}T_{m}^{ab} \,  ,
\eq$$
the RTT relation  reduces to the following commutation relation
$$
\com{T_{\ell}^{ab}}{T_m^{cd}}=\delta^{ad}T_{\ell+m}^{cb}-\delta^{cb}T_{\ell+
m}^{ad}+\lambda
     \sum_{k=0}^{\ell-1}\Bigl\{  T_{k+m}^{cb}
T_{\ell-k-1}^{ad}-T_{\ell-k-1}^{cb}
 T_{k+m}^{ad}\Bigr\}\, .
\eqlabel\mmalgebra$$
From this structure, we can define two sets of commuting operators.
One of these  is obtained by the spectral expansion of the transfer matrix
$$
  \com{I_\ell}{I_m}=0  \quad , \quad I_m\equiv \sum_{a=1}^nT_m^{aa}  \, .
\eq$$
The other set is related to the quantum determinant of the monodromy matrix
[\ref{A.G.
Izergin and V.E. Korepin, Sov. Phys. Dokl. \b{26} (1981) 653.}]
$$
 \detq{\b{T}(u)}\equiv \sum_{\sigma\in S_n }\epsilon(\sigma
)
T^{1\sigma(1)}\bigl(u-(n-1)\lambda\bigr)T^{2\sigma(2)}\bigl(u-(n-2)\lambda\b
igr)\dots
  T^{n\sigma(n)} (u)\, .
\eq$$
Here, $\sigma(i)$ is the image of $i$ under the permutation $\sigma$,
$\epsilon(\sigma)$ is the permutation's
parity and the sum is taken over all permutations of $(1\dots n)$.
The quantum determinant is analogous to the Casimir operator of a Lie
algebra in that it commutes with all generators:
$$
       \comg{\detq{\b{T}(u)}}{\b{T}(v)}=0\, .
\eq$$
This property allows one to define a second set of commuting operators
from the coefficients of the series expansion of $\detq{\b{T}(u)}$ in terms
of the
spectral parameter
$$
\com{J_m}{J_{\ell}}=0 \quad , \quad \detq{\b{T}(u)}\equiv
1+\sum_{m=0}^{\infty}u^{-(m+1)}J_m  \, .
\eq$$
A given Hamiltonian $H$ will therefore be shown to be integrable if one can
prove its symmetry under a non-trivial monodromy matrix
$$
 \com{H}{\b{T}(u)}=0 \, ,
\eq$$
which guarantees the conserved character of the involutive sets $I_m$ and
$J_m$. However, monodromy
matrices are  formidable objects which usually do not allow such
commutators to be
directly carried out.  It is therefore very useful to codify the monodromy
matrix
in a minimal form. Such a minimal coding can sometimes be realized in terms of
the yangian algebra $Y[su(n)]$. In fact, defining the lower-order yangian
generators $(Q^{ab}_0, Q^{ab}_1)$ ($a, b=1\dots n$) by
 \foot{Here and hereafter, we use the obvious  matrix notation
$(\b{T}_0\b{T}_0)^{ab}\equiv\sum_{c=1}^n T_0^{ac}T_0^{cb}$ }
$$
      Q_0^{ab}\equiv  -T^{ab}_0 \qquad , \qquad Q_1^{ab}\equiv-T_1^{ab}+
\frac{\lambda}{2}(\b{T}_0\b{T}_0)^{ab} \, ,
\eq$$
the first few of the commutation relations \mmalgebra\ read
$$\eqalign{
&\com{Q_0^{ab}}{Q_0^{cd}}=\delta^{bc}Q_0^{ad}-\delta^{da}Q_0^{cb} \cr
&\com{Q_0^{ab}}{Q_1^{cd}}=\delta^{bc}Q_1^{ad}-\delta^{da}Q_1^{cb}  \cr
&\comg{Q_0^{ab}}{\com{Q_1^{cd}}{Q_1^{ef}}}-\comg{Q_1^{ab}}{\com{Q_0^{cd}}{Q_
1^{ef}}}= \cr
&\quad\frac{\lambda^2}{4}\biggl\{
\comg{Q_0^{ab}}{\com{(\b{Q}_0\b{Q}_0)^{cd}}{(\b{Q}_0\b{Q}_0)^{ef}}}-
\comg{(\b{Q}_0\b{Q}_0)^{ab}}{\com{Q_0^{cd}}{(\b{Q}_0\b{Q}_0)^{ef}}}\biggr\}
\, .
                    }
\eq$$
These three relations define, or more precisely, completely characterize the
yangian algebra $Y[su(n)]$. The third relation is a sort of compatibility
requirement on
the different ways to reach $Q_2$ from multiple commutations involving
lower-order
charges.

One can reconstruct the whole monodromy matrix strictly from its
lower-order yangians
whenever the former possesses a trivial quantum determinant
$$
\detq{\b{T}(u)}=\hbox{c-number}\,  .
\eqlabel\cnumbercond$$
To justify the last statement, consider the following special cases of
the algebra \mmalgebra\ :
$$\eqalign{
     T_{m+1}^{ad}&= \com{T_m^{cd}}{T_1^{ac}} + \lambda (  T_{m}^{cc}
T_{0}^{ad}-T_{0}^{cc}  T_{m}^{ad}) \qquad \quad (a \not = d)  \cr
      T_{m+1}^{aa}-T_{m+1}^{cc}&= \com{T_{m}^{ca}}{T_{1}^{ac}}+ \lambda (
T_{m}^{cc}  T_{0}^{aa}-T_{0}^{cc}  T_{m}^{aa})
       \qquad\text{(no sum)}\!\!\!\!\! . }
\eqlabel\recur$$
The first of these relations allows us to compute any $T_m^{ab} (a \not=b)$
in terms of the lower-order generators  but the
second relation is not sufficient to compute the $T_m^{aa}$ by recurrence.
However, if \cnumbercond\ holds,  its spectral expansion  gives a set of
conditions on
$\sum_{a=1}^nT_m^{aa}$ which, when  supplemented by \recur , allow one to
compute all
the $T_m^{ab}$ from the lower-order yangians. This allows for
a tremendous simplification because the
symmetry of $H$ under $\b{T}(u)$ then follows from its symmetry under the
induced $Y[su(n)]$ representation
$$
\hbox{if}\, \Bigl\{\detq{\b{T}(u)}=\hbox{c-number}\Bigr\} \, \hbox{then}
\,\Bigl\{\com{H}{Q^{ab}_{(0,1)}}=0 \, \Rightarrow\,
\com{H}{\b{T}(u)}=0\Bigr\}\, .
\eq$$
When dealing with an irreducible  representation of $\b{T}(u)$, the quantum
determinant must necessarily be proportional to the identity and the
monodromy matrix can then be represented by its lower-order yangians.
However, when the considered representation is
reducible, one must exercise care  because the quantum determinant may then
be a non-trivial
operator. In the case of the Frahm-Polychronakos model, we will see  that
the symmetry algebra
is  reducible but  nevertheless possesses a trivial quantum determinant so
that in this
special case
(and for the Haldane-Shastry model), the monodromy matrix will be solely
expressed
in terms of its
${\it reducible}$ lower-order yangians.

\subsec{The yangian representation in terms of Dunkl operators}

Having discussed the theory of $su(n)$  yangians, we now focus on
the construction of specific realizations useful for long-range
interaction models. First of all, we  work in a Hilbert space of $N$
particles endowed
with
$su(n)$ spin, in which the position (momentum)  operator of particle $i$
will be denoted by ${R}_i$ ($P_i$); its spin operators are chosen to be the
$n^2$
fundamental generators ${E}_i^{ab}$  $(a,b=1\, ...\, n)$ satisfying
$$
               \com{ {E}^{ab}_i}{ {E}^{cd}_j}=\delta_{ij} \left(
\delta^{bc}{E}^{ad}_i- \delta^{ad}{E}^{cb}_i\right)\, .
\eq$$

We now define an hermitian exchange operator $\hat{K}_{ij}$, which permutes the positions of particles $i$ and $j$
$$
    \hat{K}_{ij} \ket{r_{ 1}^{ (1)}\kdots r_{ i}^{ (i)}\kdots  r_{ j}^{
(j)}\kdots  r_{ N}^{ (N)}} = \ket{r_{ 1}^{ (1)}\kdots  r_{ j}^{ (i)}\kdots
r_{ i}^{ (j)}\kdots r_{ N}^{ (N)}}\, .
\eq$$
We stress that in our notation, operator subscripts refer to particles
whereas ket subscripts (superscripts) refer to positions
(particles) so that
$$R_i\ket{r_{1}^{(1)}\kdots r_{ p}^{(i)}\kdots
 r_{ N}^{ (N)} }=r_{ p}\ket{r_{ 1}^{
(1)}\kdots r_{p}^{(i)}\kdots  r_{N}^{(N)} }.\eq$$
The  permutation operator $\hat{K}_{ij}$ satisfies
$$\eqalign{
             & \hat{K}_{ij} f ({R}_i, {P}_i) = f ({R}_j, {P}_j)
\hat{K}_{ij} \qquad\qquad \hat{K}_{ij}\hat{K}_{jk}=\hat{K}_{ik}\hat{K}_{ij}
\cr
       & \hat{K}_{ij}   f ({R}_{\ell}, {P}_{\ell}) = f ({R}_{\ell},
{P}_{\ell}) \hat{K}_{ij} \qquad\qquad
\hat{K}_{ij}\hat{K}_{k\ell}=\hat{K}_{k\ell}\hat{K}_{ij}  }\quad \left(
k,\ell \not= i,j \right)\, .
\eq$$
Here, the caret is used to stress that $\hat{K}_{ij}$
is an abstract Hilbert-space operator and therefore acts trivially on any
c-number.
This contrasts with the
$K_{ij}$ operator generally used, which exchanges the position eigenvalues
according to $K_{ij}r_i=r_jK_{ij}$ and which
  is simply  the position-space representation of this abstract operator:
$$
       \bra{r_{ 1}^{ (1)}\kdots  r_{ N}^{
(N)}}\hat{K}_{ij}\ket{\psi}=K_{ij}\bracket{r_{ 1}^{ (1)}\kdots  r_{ N}^{
(N)}}{\psi} \, .
\eq$$

In a similar way, one can also define a spin exchange operator
$$\eqalign{
             & {P}_{ij} {E}_i^{ab} = {E}_j^{ab}  {P}_{ij}  \cr
       & {P}_{ij} {E}_{\ell}^{ab} = {E}_{\ell}^{ab}  {P}_{ij}
                }\quad \left( \ell \not= i,j \right)\, .
\eq$$
In the fundamental  basis, this operator takes the simple
form $${P}_{ij}=\sum_{a,b=1}^{n}{E}^{ab}_i{E}^{ba}_j\, .\eq$$

Now, in order to eventually establish a link between spatial and
 spin models, one introduces  a projection $\Pi$ [\ref{D. Bernard, M.
Gaudin, F.D.M.
Haldane and V. Pasquier,  J. Phys. A: Math.Gen. \b{26} (1993)
5219.}\refname\gaudin],
which consists in projecting onto states that are symmetric with  respect
to the joint
interchange of position and spin variables, that is, states satisfying
${\hat{K}_{ij}}P_{ij}=1$.  In practice, this projection boils down to the
following
operation:  in a given expression, we move all   $\hat{K}_{ij}$ operators
to the
right and  replace them by ${P}_{ij}$ operators acting in reverse order;
 e.g.
$$\Pi \  \{\hat{K}_{ij}\hat{K}_{jk}\}={P}_{jk}{P}_{ij}\, .\eq$$
 This projection  possesses the following crucial properties
$$\eqalign{
      \Pi \{ A B \}&=\Pi \{ A\} \cdot \Pi \{ B\} \qquad \text{if} \quad
\com{{B}}{\hat{K}_{ij}{P}_{ij}}=0 \cr
      \Pi \{A B\}&=\Pi \{B\} \cdot \Pi \{A\} \qquad \text{if} \quad
\com{{A}}{\Pi \{B\}}=0\, .
                    }
\eqlabel\projprop$$

Using this projection technique, one can construct a spin
representation of the algebra \mmalgebra\ [\ref{M.L. Ge and Y. Wang, {\it
Long-range
interaction models and yangian symmetry, Phys. Rev. E,\/}cond-mat/9509080
(1995)}\refname\gewang]. This representation is  based on given
position-space Dunkl
operators ${\bar{D}}_i$ (and from now on, we will use the overhead bar to
indicate
that an operator acts non trivially only in position space) obeying
$$\eqalign{
      & \hat{K}_{ij}{\bar{D}}_i = {\bar{D}}_j \hat{K}_{ij}     \cr
      &\com{{\bar{D}}_i}{{\bar{D}}_j}=\lambda ({\bar{D}}_i-{\bar{D}}_j)
\hat{K}_{ij}\, .
                   }
\eq$$
By induction, one can in fact prove the more general commutation relation
$$\com{{\bar{D}}_i^{\ell}}{{\bar{D}}_j^m}=\lambda \sum_{k=0}^{\ell-1}\Bigl(
{\bar{D}}_{i}^{k+m}
{\bar{D}}_{j}^{\ell-k-1}-{\bar{D}}_i^{\ell-k-1}{\bar{D}}_j^{k+m} \Bigr)
\hat{K}_{ij}\, , \eqlabel\dunkldeux
                  $$
from which one can immediately define the following involutive set
$$
\com{\bar{ I}_{\ell}}{\bar{ I}_{m}}=0 \quad , \quad \bar{ I}_m\equiv
\sum_{i=1}^N {\bar{D}}_i^m\, .
\eq$$
These quantities are purely spatial; in order to define spin invariants,
we can use the properties \projprop\ and \dunkldeux\  to show that the currents
$$
        T_m^{ab}=\sum_{i=1}^N {E}_i^{ba}\Pi \bigl\{  {\bar{D}}_i^m\bigr\}
\eqlabel\gewang$$
satisfy the monodromy matrix algebra \mmalgebra . The involutive ${I}_m$
set associated with this algebra is then simply given by
$$
  \com{{I}_m}{{I}_{\ell}}=0 \quad , \quad {I}_m= \Pi\bigl\{ \bar{
I}_m\bigr\} =\sum_{i=1}^N\Pi \bigl\{ {\bar{D}}_i^m \bigr\}.
\eq$$
Quite remarkably, the  monodromy matrix \gewang\ can also be expressed
in the form  $\b{T}(u) =\Pi\{ \b{T'}(u) \}$,
with $\b{T'}(u)$  another representation of the algebra \mmalgebra , given
by [\gaudin]
$$
    \b{T'}(u) = \frac{1}{\bar{\Delta} (u)}\prod_{i=1}^N\Bigl\{
(u-\bar{D}'_i)\b{1}_i + \lambda \b{E}_i^{\top}   \Bigr\}
 \quad , \quad
        \bar{\Delta}(u) \equiv \prod_{i=1}^N (u-\bar{D}'_i)
\eq$$
where the spin operators have been grouped in matrix form according
to  $\bigl(\b{E}_i^{\top}\bigr)^{ab}=E_i^{ba}$
and the product over $i$ includes both the usual linear matrix product as
well as the tensor one.
Here, the $\bar{D}'_i$ are modified  Dunkl operators
$$
        \bar{D}'_i\equiv {\bar{D}}_i-\lambda \sum_{\st j=1 \atop\st  j <
i}^N \hat{K}_{ij}  \, ,
\eq$$
which satisfy the degenerate affine Hecke algebra with respect to
position-space permutations
$$\eqalign{
 & \hat{K}_{i i\pm 1}\bar{D}'_i - \bar{D}'_{i \pm 1}\hat{K}_{i i\pm 1}=\pm
\lambda \cr
&\com{\bar{D}'_i}{\bar{D}'_j}=0 \, .
  }
\eqlabel\moddunklcom$$
Using this Hecke algebra, one can show that
$$\com{\bar{\Delta}(u)}{\hat{K}_{ii+1}}=\com{\bar{\Delta}(u)}{\hat{K}_{ii+1}
P_{ii+1}}=0\, ,
\eq$$
which can in turn be used to prove
that$\com{\b{T'}(u)}{\hat{K}_{ii+1}P_{ii+1}}=0$.
But since any permutation  can be expressed as a product of transpositions,
we actually have
$$
\com{\bar{\Delta}(u)}{\hat{K}_{ij}}=\com{\bar{\Delta}(u)}{\hat{K}_{ij}{P}_{i
j}}=\com{\b{T'}(u)}{\hat{K}_{ij}{P}_{ij}}=0 \, .
\eqlabel\deltaprop$$
Now the quantum determinant of $\b{T'}(u)$ having already been calculated  as
$$
   \detq{\b{T'}(u)}=\frac{  \bar{\Delta}(u+\lambda)}{  \bar{\Delta}(u)}\, ,
\eq$$
one can use the property  \deltaprop\ to factorize the projection and calculate
the quantum determinant of $\b{T}(u)$ as [\gaudin]
$$
\detq{\b{T}(u)}=\detq{\Pi \{\b{T'}(u) \}}=\Pi \Bigl\{ \detq{\b{T'}(u)}
\Bigr\}=\Pi
\Biggl\{\frac{\bar{\Delta}(u+\lambda)}{\bar{\Delta}(u)}\Biggr\}\, .
\eqlabel\detqgewang$$
In principle, one can extract the $\{J_k\}$ set from this formula but the
result
is highly cumbersome. It is much
simpler to focus instead on $\Delta (u)$. Indeed, the relations
\moddunklcom\ and
\deltaprop\  also imply
$$
     \com{\bar{\Delta} (u)}{\bar{\Delta} (v)}=\com{ \Delta (u)}{ \Delta
(v)}=0 \, ,
\eq$$
where $\Delta (u)\equiv \Pi\{\bar{\Delta}(u)\}$.
One can therefore  define a simple set of commuting operators by using
$\frac{\d}{\d u}\ln [\Delta (u)]$
as their generating function [\talhal]
$$\eqalign{
      &\com{\bar{H}_m}{\bar{H}_{\ell}}=0 \quad , \quad \bar{H}_m\equiv
\sum_{i=1}^N  \bigl( \bar{D}'_i\bigr)^m\cr
      &\com{H_m}{H_{\ell}}=0 \quad , \quad H_m\equiv \Pi\bigl\{\bar{H}_m
\bigr\}=\sum_{i=1}^N\Pi\Bigl\{\bigl(\bar{D}'_i\bigr)^m\Bigr\}\, .
}
\eq$$
By virtue of \detqgewang , this new set is obviously equivalent to the
$\{J_m\}$ set and from now on, we will focus on the sets $\{I_m\}$ and
$\{H_m\}$.

\newsec{The Dynamical Calogero Model}

In order to obtain the Hamiltonian spin-chain conservation laws, one must first
consider an $N$-body dynamical Calogero model in which the
particles are chosen to have unit mass and are allowed to move along the
line,  under the influence of a position-space exchange
interaction and subject to  an harmonic confinement
$$
     \bar{\cal H}^{\hbox{\scriptsize (CC)}}\equiv
\frac{1}{2}\sum_{i=1}^{N}P_i^2-\frac{1}{2}\sump{i,
j=1}{N}\frac{g\bigl(g-\hbar \hat{K}_{ij}
\bigr)}{R_{ij}R_{ji}}+\frac{1}{2}\omega^2\sum_{i=1}^{N} R_i^2 \, .
\eq$$
The integrability of this model has been demonstrated e.g., in  [\ref{A. P.
Polychronakos, Nucl. Phys. \b{B419}
(1994) 553.}\refname\polyspectrum] by  means of the  operators
$$
         {\bar{\cal D}}_j^{\pm}\equiv P_j+\sqrt{-1}\, g\sum_{\st k=1\atop\st
k\not=j}^N\frac{1}{R_{jk}}\hat{K}_{jk}\pm \sqrt{-1}\, \omega R_j    \, ,
\eqlabel\dpm$$
which satisfy $\bigl(\bar{\cal D}_j^{\pm}\bigr)^{\dagger}=\bar{\cal
D}_j^{\mp}$, in addition to the commutation properties
$$   \eqalign{
      & \com{{\bar{\cal D}}_i^{\pm}}{{\bar{\cal D}}_j^{\pm}}=0 \cr
      & \com{{\bar{\cal D}}_i^{\pm}}{{\bar{\cal D}}_j^{\mp}}=
            \mp 2\omega\delta_{ij}\biggl( \hbar+g\sum_{\st k=1\atop\st
k\not= i}^N\hat{K}_{ik} \biggr)\pm (1-\delta_{ij})2\omega g \hat{K}_{ij}\,
.
}
\eqlabel\calodcom$$
From these,  one can define [\ref{K. Hikami, {\it Yangian symmetry and Virasoro
 character in a lattice spin system with long-range interactions, Nucl.
Phys.\/}
\b{B441} [FS] (1995) 530.}\refname\hikami] the Dunkl operator
${\bar{\cal D}}_i\equiv {\bar{\cal D}}_i^{+}{\bar{\cal D}}_i^{-}+\hbar\omega$
$$\eqalign{
        {\bar{\cal D}}_i=-g^2\!\!\!\!\!\sum_{\st j,k=1 \atop \st
i\not=j\not=k\not=i}^N\!\!\!
            \frac{1}{R_{jk}R_{ki}}\hat{K}_{ijk}
+\sum_{\st j=1\atop\st j\not=i}^N\biggl\{ \sqrt{-1}\,  &g
\frac{1}{R_{ij}}\bigl( P_i+P_j \bigr)
-\omega g  \biggr\}\hat{K}_{ij} \cr
       &+P_i^2-\sum_{\st j=1\atop\st j\not=i}^N\frac{g(g-\hbar
\hat{K}_{ij})}{R_{ij}R_{ji}}+\omega^2 R_i^2  \cr
}
\eqlabel\dunkldyn$$
(where here and hereafter, we use the notation $\hat{K}_{i_1\dots
i_n}\equiv \prod_{j=1}^{n-1}\hat{K}_{i_ji_{j+1}}$).  The  deformation
parameter of this Dunkl operator is  $\lambda=-2\omega g$; it therefore
satisfies the commutation relation:
$$
       \com{ {\bar{\cal D}}_i}{ {\bar{\cal D}}_j}=- 2\omega g \bigl(
{\bar{\cal D}}_i-{\bar{\cal D}}_j\bigr)\hat{K}_{ij}\, .
\eqlabel\dunkldyncom$$
The representation of the yangian algebra induced by this Dunkl operator is
given by
$$\eqalign{
  & \b{{\cal Q}_0}=-\sum_{i=1}^N\b{E}_i^{\top}  \cr
 & \b{{\cal
Q}_1}=g^2\sump{i,j,k=1}{N}(\b{E}_i\b{E}_j\b{E}_k)^{\top}\frac{1}{R_{ij}R_{jk
}}+\sump{i,j=1}{N}(\b{E}_i\b{E}_j)^{\top}
     \biggl\{ \frac{\hbar g}{R_{ij}^2}-\frac{\sqrt{-1}\,
g}{R_{ij}}\bigl(P_i+P_j \bigr)
\biggr\} \cr
&\qquad\qquad\qquad\qquad\qquad\qquad{}-\sum_{i=1}^N\b{E}_i^{\top}\Biggl\{
P_i^2+g^2\sum_{\st j=1\atop\st j\not=i}^N\frac{1}{R_{ij}^2}+\omega^2
R_i^2\Biggr\}-N\omega g\b{1}\, .
}
\eqlabel\qyangdyn$$
Its associated monodromy matrix then allows us to generate two non-trivial
involutive sets of operators, denoted $\bar{{\cal  I}}_m$ and $\bar{{\cal
H}}_m$ (calligraphic
symbols being used for the charges pertaining to the dynamical model).
Their first
member is given explicitly by
$$\eqalignno{
       &\bar{\cal  I}_1=\sum_{i=1}^N \bar{\cal D}_i=2\bar{\cal
H}^{\hbox{\scriptsize (CC)}}-\omega g
\sum_{i,j=1}^N\hat{K}_{ij}\eqnumlabel\tata\cr
       &\bar{\cal  H}_1=\sum_{i=1}^N\biggl( {\bar{\cal D}}_i+2\omega
g\sum_{\st j=1 \atop \st j<i}^N\hat{K}_{ij}\biggr)=2\bar{\cal
H}^{\hbox{\scriptsize
(CC)}}\, .\eqnumlabel\toto  }
$$
Because $\bar{ \cal H}^{\hbox{\scriptsize (CC)}}$ is included in $\{\bar{
\cal H}_m\}$ and these essentially arise from an expansion of
the quantum determinant, the symmetry of $\bar{ \cal H}^{\hbox{\scriptsize
(CC)}}$ under the monodromy matrix induced by this Dunkl operator is
manifest. This means that both the $\bar{\cal I}_m$ and $\bar{\cal H}_m$
sets constitute involutive invariants for this dynamical model.
Moreover, the  basic relations \calodcom\ can be shown [\polyspectrum] to imply
$$
        \com{ \bar{\cal H}^{\hbox{\scriptsize (CC)}}}{({\bar{\cal
D}}_k^{\pm})^n}=\pm n\hbar\omega ({\bar{\cal D}}_k^{\pm})^n \, ,
\eqlabel\creatdyn$$
thereby furnishing a set of creation operators from  which  the spectrum
can be readily obtained.

\newsec{The Frahm-Polychronakos spin chain}

The Frahm-Polychronakos spin-chain model is defined by \fp. In this
expression, the $x_i$'s are the zeroes of the Hermite polynomials.  In this
section, we will consider a generalized version of the FP Hamiltonian
$H(\b{r})$:
$$
           H(\b{r})=\frac{1}{2}\sump{i, j=1}{N}\frac{1}{r_{ij}r_{ji}}{P}_{ij} \, .
\eqlabel\fpgeneral$$
 in which $\b{r}$ is a set of unconstrained position eigenvalues, in order
to see explicitly how the yangian symmetry picks up the particular FP
model, i.e.  how it enforces $r_i=x_i$.  In the following, the (potential)
conserved charges pertaining to this general version of the spin chain that
are inherited from the dynamical model will be denoted by ${\cal I}_m^{
\hbox{\scriptsize (0)}}(\b{r})$ and
${\cal H}_m^{ \hbox{\scriptsize (0)}}(\b{r})$. The subindex $0$ refers to
an $\hbar$-expansion to be explained shortly.

In order to generate a candidate symmetry algebra for this generalized model, we
consider the  position-space representation of the dynamical Calogero
model, in which,  from now on, we set $\omega=g=1$.
The spin part of $\bar{\cal H}^{\hbox{\scriptsize (CC)}}$ is simply
isolated as the linear $\hbar$-piece of the Hamiltonian.  More generally,
the spin part is obtained by differentiating the Hamiltonian with respect
to $\hbar$ and, since the kinetic term is quadratic in $\hbar$, setting
$\hbar=0$ at the end. This ignores the fact that the zeroes should be fixed
at particular positions, but nevertheless suggests to consider the
$\hbar$-expansion of the Dunkl operators and the related conserved
operators.

Let us then expand the dynamical Dunkl operator \dunkldyn\ according to
$\bar{\cal D}_i=\sum_{k}\bar{\cal D}_i^{\hbox{\scriptsize
(}k\hbox{\scriptsize )}}\hbar^k$ and for the time being, concentrate on the
zeroth order term:
$$
 \bar{{\cal D}}_i^{ (0)}(\b{r})=-\!\!\!\!\!\sum_{\st j,k=1 \atop \st
i\not=j\not=k\not= i}^N\!\!\!\frac{1}{r_{jk}r_{ki}}K_{ijk}  -\sum_{\st
j=1\atop\st j\not=i}^NK_{ij}
       +\sum_{\st j=1 \atop \st j\not=i}^N\frac{1}{r_{ij}^2}+r_i^2 \, .
\eq$$
Since the Dunkl algebra \dunkldyncom\ is satisfied for all values of
$\hbar$, $\bar{\cal D}_i^{ \hbox{\scriptsize (0)}}(\b{r})$
is also a genuine Dunkl operator, with deformation parameter $\lambda =
-2$. Now for generic values of the $r_i$'s,
the induced $Y[su(n)]$ representation is irreducible and its quantum
determinant is therefore a trivial c-number.
As a corollary, the  $\{{\cal H}_m^{ (0)}(\b{r})\}$ do not provide
non-trivial conserved charges, i.e. these quantities are independent
of any  exchange operators. On the other hand, the set $\{{\cal I}_n^{
\hbox{\scriptsize (0)}}(\b{r})\}$ does provide a non-trivial involutive
ensemble, its first member being given by
$$
       {\cal I}_1^{(0)}(\b{r})=\sum_{i=1}^{N}\Pi \Bigl\{\bar{{\cal
D}}_i^{(0)}(\b{r})\Bigr\}=-\sump{i,j=1}{N}{P}_{ij}
+\sump{i,j=1}{N}\frac{1}{r_{ij}^2}+\sum_{i=1}^Nr_i^2 \, .
\eq$$
To obtain this result, we used the identity
$$
\sump{j,k,\ell=1}{N}\,\frac{1}{r_{jk}r_{k\ell}}=\frac{1}{3}\sump{j,k,\ell=1}
{N}\biggl\{  \frac{1}{r_{jk}r_{k\ell}}+\frac{1}{r_{k\ell}r_{\ell
j}}+\frac{1}{r_{\ell j}r_{jk}}\biggr\}\equiv 0\, .
\eqlabel\ident$$
Now, the higher order ${\cal I}_m^{ \hbox{\scriptsize (0)}}(\b{r})$ do not
contain any term having the form of $H(\b{r})$.  In other words, this set
of commuting operators has  no relation at this point with the generalized
model defined by the Hamiltonian $H(\b{r})$.  In order for the  set
$\{{\cal I}_m^{ \hbox{\scriptsize (0)}}(\b{r})\}$ to represent involutive
invariants for \fpgeneral , one must enforce the invariance of  $H(\b{r})$
under the corresponding $Y[su(n)]$ algebra, whose lower-order generators
are
given by
$$\eqalign{
  & \b{{\cal Q}}_{\b{0}}^{ \hbox{\scriptsize
(0)}}(\b{r})=-\sum_{i=1}^N\b{E}^{\top}_i   \cr
 & \b{{\cal Q}}_{\b{1}}^{ \hbox{\scriptsize
(0)}}(\b{r})=\sump{i,j,k=1}{N}(\b{E}_i\b{E}_j\b{E}_k)^{\top}\frac{1}{r_{ij}r
_{jk}}-\sum_{i=1}^{N}\b{E}_i^{\top}\Biggl\{
    \, \sum_{\st j=1\atop\st j\not=i}^N\frac{1}{r_{ij}^2}+r_i^2
\Biggr\}-N\b{1} \, .
}
\eqlabel\yangfp$$
In other words, we require that $\com{H(\b{r})}{\b{{\cal Q}}_{\b{(0,1)}}^{
\hbox{\scriptsize (0)}}(\b{r})}=0$.  A direct calculation [\hikami] shows
that this holds if and only if the variables $r_i$ obey
$$
\sum_{\st j=1\atop\st j\not=i}^N\frac{1}{r_{ij}^3}=\frac{1}{2}r_i \, .
\eqlabel\sumpube$$
One can show that this condition is satisfied by the  zeroes (written
$x_i$) of the
Hermite polynomial $H_N(x)$ (cf. Appendix B). In fact, by judiciously
substracting known summation identities  [\ref{F. Calogero, Lett. Nuovo.
Cim. \b{20} (1977)
 14; S. Ahmed, M. Bruschi and F. Calogero, Lett. Nuovo. Cim. \b{49 B}
(1979) 2.}\refname\zeroesi]  for these numbers, one can
generate a whole sequence of `higher order' identities, the simplest of
them being  listed in  Appendix B; these will play a crucial role in
subsequent calculations.

In retrospect, by freezing the positions of the particles on the zeroes of
$H_N(x)$, we send $H(\b{r})$ on $H^{\hbox{\scriptsize (FP)}}$ and thus
obtain an integrable $Y[su(n)]$-symmetric spin chain with a non-trivial
involutive set of invariants given by
$$
   { I}_m(\b{x})\equiv\lim_{\b{r} \rightarrow \b{x}} {\cal I}_m^{
\hbox{\scriptsize (0)}}(\b{r})=\sum_{i=1}^N \Pi\biggl\{\Bigl(
                 \bar{\cal D}_i^{ \hbox{\scriptsize (0)}}(\b{x}) \Bigr)^m
\biggr\} \, .
\eqlabel\infp$$
These are the conserved quantities first found by Polychronakos [\polylat]
(but without the yangian interpretation).

Moreover, defining ${\bf\cal C}^{\bf\pm}_m\equiv
\sum_{i=1}^N\b{E}_i\bigl(\bar{{\cal D}}_i^{ \pm}\bigr)^m$, expanding
\creatdyn\ to ${\cal O}(\hbar)$ and setting $\omega=g=1$, we find
$$
     \com{\bar{\cal H}^{\hbox{\scriptsize (CC)(0)}}(\b{r})}{{\bf\cal
C}^{\pm\hbox{\scriptsize (1)}}_m(\b{r})}+\com{\bar{\cal
H}^{\hbox{\scriptsize (CC)(1)}}(\b{r})}{{\bf\cal C}^{\pm\hbox{\scriptsize
(0)}}_m(\b{r})}=\pm m \,{\bf\cal C}^{\pm\hbox{\scriptsize (0)}}_m(\b{r}) 
\, .
\eqlabel\creatcom$$
Because $\bar{\cal H}^{\hbox{\scriptsize (CC)(0)}}(\b{r})$ is scalar and
$K_{ij}$-invariant, the first commutator on the left hand side  reduces to
the action of the derivative on $\bar{\cal H}^{\hbox{\scriptsize
(CC)(0)}}(\b{r})$, which is given by
$$
      \frac{\d}{\d r_k}\Bigl\{\bar{\cal H}^{\hbox{\scriptsize
(CC)(0)}}(\b{r})\Bigr\}=\frac{1}{2}\frac{\d}{\d r_k}\Biggl\{
\sump{i,j=1}{N}\frac{1}{r_{ij}^2}+\sum_{i=1}^{N}r_i^2 \Biggr\} =
r_k-2\sum_{\st j=1\atop\st j\not=k}^{N}
      \frac{1}{r_{jk}^3}  \, .
\eqlabel\temp$$
This vanishes as  $\b{r}\rightarrow \b{x}$ (cf. the identity (B.7)) and
\creatcom\ takes the form
$$
   \comg{\frac{1}{2}\sump{i,j=1}{N}\frac{1}{x_{ij}x_{ji}}K_{ij}}{{\bf\cal
C}^{\pm\hbox{\scriptsize (0)}}_m(\b{x})}=\pm m\,  {\bf \cal
C}^{\pm\hbox{\scriptsize (0)}}_m(\b{x}) \, .
\eqlabel\form$$
Taking now the projection and using the $K_{ij}P_{ij}$-invariance of the
two commuted  operators,
we obtain a whole set of creation operators for the FP model:
$$
  \com{H^{\hbox{\scriptsize (FP)}}}{\b{C}^{\pm}_{m}}=\pm m\,  \b{C}^{\pm}_m
\quad , \quad \b{C}^{\pm}_m
        \equiv \Pi\Biggl\{\sum_{k=1}^{N}\b{E}_k \sum_{\st \ell=1\atop\st
\ell\not=k}^{N}\biggl(\frac{1}{x_{k\ell}}K_{k\ell}\pm  x_k\biggr)^m\Biggr\}
\, .
\eqlabel\creatfp$$
These generalize the lower-order creation operators found in
[\hikami,\frahm]. We therefore possess a set of non-trivial creation
operators $\b{C}^{\pm}_m$
and conservation laws $I_m$.  However, as previously pointed out, the
$\{{\cal H}^{ \hbox{\scriptsize (0)}}_m\}$ set associated with the symmetry
algebra is trivial. Since  $\{{\cal I}^{ \hbox{\scriptsize (0)}}_m\}$ does
not contain the defining
Hamiltonian, a whole set of commuting conservation laws is still missing.

\newsec{The Hamiltonian conservation laws of the FP model}

Our proof for the commutativity of the conservation laws will strongly rely
on the structure of the FP Hilbert
space. For the $su(2)$ Haldane-Shastry model, the yangian symmetry algebra
has been shown to be a direct sum
of irreducible $Y[su(2)]$ ``motif" representations, each possible motif
appearing with unit multiplicity [\gaudin] . This result has been obtained
by calculating the dimensions of the $Y[su(2)]$ motif representations as a
tensor product
of $su(2)$ spin representations and then showing that these motifs exhaust
the Hilbert space. For the $su(n)$ case ($n>2 $), the motifs are not
expressible as a  free tensor product [\hikami]  and to our knowledge,  it
hasn't been proved that the $Y[su(n)]$ motifs exhaust the Hilbert space.
However, strong numerical evidence [\hikami] suggests that the symmetry
algebras for both the $su(n)$ HS and FP models are  also a direct sum of
${\it non-degenerate}$ motifs. In the following, we will consider this
statement to be true.

The non-degenerate character of the motifs implies that any  two operators
$A$ and  $B$ commuting with the monodromy matrix $\b{T}(u)$ must also
commute amongst themselves (see e.g., [\talhal]). Indeed,  the Hilbert
space of our reducible $Y[su(n)]$ invariant theory  contains a certain
number of yangian highest-weight states, each of which is associated with a
given motif.  These highest-weight states are  eigenvectors of the
diagonal elements $T^{aa}(u)$ ($a=1\dots n$), with eigenvalues that
completely specify the given motif. Since the motifs have unit
multiplicity, the highest-weight states $T^{aa}(u)$-eigenvalues  form
non-degenerate sets. Now consider the two states $ AB \ket{\Lambda}$ and $
BA \ket{\Lambda}$, where $ \ket{\Lambda}$ is a yangian highest-weight
state. Since $A$ and $B$ commute with $\b{T}(u)$, both these states will be
eigenvectors of $T^{aa}(u)$ with the same eigenvalue. But since these
eigenvalues are non-degenerate, the two states must in fact  be
proportional to one another, which implies $\com{A}{B}=0$ on any
highest-weight state.
In fact, since all of the states can be generated by acting on the
highest-weight states with lowering
operators of the form $\prod_{i}T^{a_ib_i}(\lambda_i)$ (with $a_i<b_i$ and
the $\lambda_i$ chosen
to satisfy a set of Bethe ansatz equations),  one sees that $A$ and $B$
will in fact
commute in the entire Hilbert space.

We will now prove that the first order terms in the $\hbar$-expansion of
the dynamical $\{{\cal H}_m\}$ set satisfy
the FP $Y[su(n)]$ symmetry and are therefore in involution. Starting from
the dynamical symmetry
$\com{\b{\cal Q}_{ \b{(0,1)}}(\b{r})}{{\cal H}_m(\b{r})}=0$ and expanding
to ${ \cal O}(\hbar)$, we have
$$
  \com{\b{{\cal Q}}_{ \b{(0,1)}}^{  (0)}(\b{r})}{{\cal H}{}_m^{ (1)}(\b{r})}
     =-\com{\b{{\cal Q}}_{\b{(0,1)}}^{ (1)}(\b{r})}{{\cal H}_m^{
(0)}(\b{r})}\, .
\eqlabel\freeze$$
On the right hand side, the commutation with $\b{{\cal Q}}_{\b{0}}^{ (1)}$
is trivially zero, while that with $\b{{\cal Q}}_{\b{1}}^{ (1)}$
can be greatly simplified by appealing to the scalar nature of ${\cal
H}_m^{ (0)}(\b{r})$ and its invariance under $K_{ij}$ and $K_{ij}P_{ij}$:
$$\com{\b{{\cal Q}}_{\b{1}}^{  (0)}(\b{r})}{{\cal H}_m^{  (1)}(\b{r})}=
\sump{i,j=1}{N}(\b{E}_i\b{E}_j)^{\top}\frac{1}{r_{ij}}\biggl\{\frac{\d}{\d
r_i}\Bigl( {\cal H}_m^{  (0)}(\b{r})\Bigr)+\frac{\d}{\d r_j}\Bigl( {\cal
H}_m^{  (0)}(\b{r})\Bigr)  \biggr\}\, .
\eqlabel\ysunsyma$$
To further simplify the right hand side, we will now explicitly calculate
${\cal H}_m^{  (0)}(\b{r})$. To this end, let us return to the abstract
Hilbert-space formalism and consider the following integral
$$\eqalign{
        F_m(\b{r})&\equiv \int dy_{ 1} \dots dy_{ N}\bra{y_{ 1}^{
(1)}\kdots y_{ N}^{ (N)}}{\cal H}_m^{
 (0)}(R_1\dots R_N)\sym \, , \cr
  }
\eqlabel\fr$$
where $\sym\equiv \sum_{\sigma \in S_N}\ket{r_{\sigma (1)}^{  (1)}\ldots
r_{\sigma (N )}^{(N)}}\!\!\!$  and the notation
${\cal H}_m^{  (0)}(R_1\dots R_N)$ is used to stress that these charges are
$K_{ij}$-independent. Applying
${\cal H}_m^{  (0)}(R_1\dots R_N)$  to the left  yields
$$\eqalign{
      F_m(\b{r})&=\sum_{\sigma\in S_N}\int dy_{ 1}\dots dy_{ N}\,  {\cal
H}_m^{  (0)}(y_{ 1}\dots y_{ N})
\bracket{y_{ 1}^{  (1)}\kdots y_{ N}^{ (N)}}
        {r_{\sigma (1)}^{  (1)}\ldots r_{\sigma (N )}^{(N)}} \cr
       &=\sum_{\sigma\in S_N}\int dy_{ 1}\dots dy_{ N}\,  {\cal H}_m^{
(0)}(y_{ 1}\dots
y_{ N})\delta(y_{ 1}-r_{\sigma (1)})\dots \delta(y_{ N}-r_{\sigma(N)}) \cr
       &=\sum_{\sigma\in S_N} {\cal H}_m^{  (0)}(r_{\sigma (1)}\dots
r_{\sigma (N)})=N!\  {\cal H}_m^{  (0)}(r_{ 1}\dots r_{ N}) \, ,
}
\eqlabel\rais$$
where we have used the $K_{ij}$-invariance of  ${\cal H}_m^{ (0)}(r_{ 1}\dots r_{ N})$ in the last step. We therefore have the
following result
$$
 {\cal H}_m^{ (0)}(\b{r})=\frac{1}{N!}F_m(\b{r})\, .
\eqlabel\hf$$
On the other hand, going back to \fr , one can express $F_m(\b{r})$ in the form
$$\eqalign{
      &F_m(\b{r})= \int dy_{ 1} \dots dy_{ N}\bra{y_{ 1}^{  (1)}\kdots y_{
N}^{ (N)}}\sum_{i=1}^N\Bigl\{ {\bar{\cal D}_i}'^{ \hbox{\scriptsize
(0)}}\Bigr\}^m\sym \, ,
 }
\eqlabel\second$$
where the modified Dunkl operator has the  following explicit expression
$$
 {\bar{\cal  D}_i}'^{ \hbox{\scriptsize (0)}}=\bar{\cal D}_i^{
\hbox{\scriptsize (0)}}+2\sum_{\st j=1\atop\st j<i}^N\hat{K}_{ij}=
\!\!\!\!\!\sum_{\st j,k=1\atop\st i\not=j\not=k\not=
i}^N\!\!\!\frac{-1}{R_{jk}R_{ki}}\hat{K}_{ijk}  +\sum_{\st j=1\atop\st
j\not=i}^N\sgn{i-j}\hat{K}_{ij} +\sum_{\st j=1\atop\st
j\not=i}^N\frac{1}{R_{ij}^2}+R_i^2   \,  .
\eqlabel\modunkfp$$
Applying now ${\cal H}_m^{ (0)}(R_1\dots R_N)$ to the right and  using
$$\eqalignno{
   &\hat{K}_{ij}\sym=\sym \eqnumlabel\syme\cr
   &f(R_i)\sym=\sum_{\sigma\in S_N}f(r_{\sigma (i)})\ket{r_{\sigma (1)}^{
(1)}\ldots r_{\sigma (N )}^{(N)}}\, , \eqnumlabel\frik }
$$
one obtains
$$\eqalign{
F_m(\b{r})= \sum_{\sigma\in S_N}\sum_{i=1}^N\Biggl\{ \sum_{\st
j,k=1\atop\st i\not=j\not=k\not=i}^N\!\!\!&\frac{-1}{r_{\sigma{ (j)}\sigma
(k)}r_{\sigma{ (k)}\sigma (i)}}\cr
+&\sum_{\st j=1\atop\st j\not=i}^N\sgn{i-j}+ \sum_{\st j=1\atop\st
j\not=i}^N\frac{1}{r_{\sigma{ (i)}\sigma (j)}^2}+r_{\sigma (i)}^2
\Biggr\}^m\, .
}\eqlabel\inter$$
Considering now the $ (N-1)!$ permutations for which $\sigma(i)=\ell$, this
can be  rewritten  as
$$
    F_m(\b{r})= (N-1)!\sum_{\ell=1}^N\sum_{i=1}^N\Biggl\{ \sum_{\st
j,k=1\atop\st
\ell\not=j\not=k\not=\ell}^N\!\!\!\frac{-1}{r_{jk}r_{k\ell}}+\sum_{\st
j=1\atop\st j\not=i}^N\sgn{i-j}+
\sum_{\st j=1\atop\st j\not=\ell}^N\frac{1}{r_{j\ell}^2}+r_{\ell}^2
\Biggr\}^m\, .
\eqlabel\almost$$
Finally, using a binomial expansion to factorize the $\sgn{i-j}$ term
(i.e., the $r_i$-independent piece) and using \hf , we find
$$
   {\cal H}_m^{ (0)}(\b{r})=\sum_{\ell=1}^N\sum_{p=0}^mC_{pm}\Biggl\{
\sum_{\st j,k=1\atop\st
\ell\not=j\not=k\not=\ell}^N\!\!\!\frac{-1}{r_{jk}r_{k\ell}}+\sum_{\st
j=1\atop\st j\not=\ell}^N\frac{1}{r_{j\ell}^2}+r_{\ell}^2 \Biggr\}^p\, ,
\eqlabel\result$$ where
$$C_{pm}\equiv \frac{1}{N}\pmatrix{m \cr
p}\sum_{i=1}^N\Bigl\{2i-(N+1)\Bigr\}^{m-p}\, .
\eq$$

To complete the calculation of the commutator \freeze , we need to
 evaluate the action of the derivative on ${\cal H}_m^{ (0)}(\b{r})$ (cf.
\ysunsyma ) and freeze the particle positions:
$$\eqalign{
   \lim_{\b{r}\rightarrow \b{x}}\frac{\d}{\d r_i}\biggl\{{\cal H}_m^{ (0)}
(\b{r})\biggr\}=
\lim_{\b{r}\rightarrow \b{x}}\sum_{\ell=1}^N\sum_{p=0}^mC_{pm}\,
 p&\Biggl\{  \sum_{\st j,k=1\atop\st
\ell\not=j\not=k\not=\ell}^N\!\!\!\frac{-1}{x_{jk}x_{k\ell}}+\sum_{\st
j=1\atop\st
j\not=\ell}^N\frac{1}{x_{j\ell}^2}+x_{\ell}^2 \Biggr\}^{p-1} \cr
&\cdot\frac{\d}{\d r_i} \Biggl\{   \sum_{\st j,k=1\atop\st
\ell\not=j\not=k\not=\ell}^N\!\!\!\frac{-1}{r_{jk}r_{k\ell}}+\sum_{\st
j=1\atop\st
j\not=\ell}^N\frac{1}{r_{j\ell}^2}+r_{\ell}^2 \Biggr\}\, . }
\eqlabel\derive$$
In such calculations, the $\b{r}\rightarrow \b{x}$ limit is not a simple
substitution and must be taken with care.
Indeed, the summation formulae for $x_i$ are valid for {\it numbers\/} and
are therefore not preserved by the action of the derivatives.
This means that one may take the substitution $ \b{r}\rightarrow \b{x}$ and
use the simplifying identities only
if the targeted expression is no longer acted upon by any derivatives. In
light of this remark, we see that we cannot simplify
the second bracketed factor in \derive\ without first carrying out
the differentiation.  However, we can use the formulae (B.7), (B.8) and
(B.11) to reduce the first bracketed factor right away:
$$\eqalign{
\sum_{\st k=1\atop\st k\not=\ell}^N\!\!\!\frac{1}{x_{\ell
k}}\biggl(-\sum_{\st j=1\atop\st
j\not=k}^N\!\!\!\frac{1}{x_{k j}} + \frac{1}{x_{k\ell}}\biggr)+\sum_{\st
j=1\atop\st
j\not=\ell}^N\frac{1}{x_{j\ell}^2}+x_{\ell}^2&= -\sum_{\st k=1\atop\st
k\not=\ell}^N\!\!\!\frac{x_k}{x_{\ell k}} +x_{\ell}^2 \cr
&=-x_{\ell}^2+(N-1)+x_{\ell}^2=N-1\, .}
\eq$$
We are thus left with
$$\eqalign{
   \lim_{\b{r}\rightarrow \b{x}}&\frac{\d}{\d r_i}\biggl\{{\cal H}_m^{
(0)}(\b{r})\biggr\}=\cr
&\lim_{\b{r}\rightarrow \b{x}}\sum_{\ell=1}^N\sum_{p=0}^mC_{pm}\,
p(N-1)^{p-1}\frac{\d}{\d r_i} \Biggl\{   \sum_{\st j,k=1\atop\st
\ell\not=j\not=k\not=\ell}^N\!\!\!\frac{-1}{r_{jk}r_{k\ell}}+\sum_{\st
j=1\atop\st j\not=\ell}^N\frac{1}{r_{j\ell}^2}+r_{\ell}^2 \Biggr\}\, .
}\eqlabel\simplefp$$
Commuting the sum over $\ell$ past the derivative and using the identity \ident\
 we finally obtain
$$\eqalign{
    \lim_{\b{r}\rightarrow \b{x}}\frac{\d}{\d r_i}\biggl\{{\cal H}_m^{
(0)}(\b{r})\biggr\}&= \lim_{\b{r}\rightarrow \b{x}}
    \sum_{p=0}^mC_{pm}\, p(N-1)^{p-1}\frac{\d}{\d r_i} \Biggl\{ \sum_{\st
\ell=1\atop\st  \ell\not=i}^N\frac{1}{r_{\ell i}^2}
    +  \sum_{\st j=1\atop\st  j\not=i}^N\frac{1}{r_{ij}^2}+r_i^2 \Biggr\} \cr
    &=\sum_{p=0}^mC_{pm}\, p(N-1)^{p-1}2\biggl\{   x_i-2\sum_{\st j=1
\atop\st j\not=i}^N\frac{1}{x_{ij}^3}\biggr\}=0\, ,
}
\eqlabel\zero$$
where we have used  relation (B.9) in the last step. Substituting this
result back in \ysunsyma , we see that
the quantities $H_{2m}\equiv \lim_{\b{r}\rightarrow \b{x}}{\cal H}_m^{
(1)}(\b{r})$ satisfy the FP $Y[su(n)]$ symmetry.  In other words, the
 $H_{2m}$ all commute with the monodromy matrix.  As already pointed out,
this implies that
they are necessarily in involution.

A compact but implicit expression for the $H_{2m}$ following from the
$\hbar$-expansion of the dynamical operators, is given by
$$
  H_{2m}=\lim_{\b{r}\rightarrow \b{x}}\sum_{i=1}^N
\sum_{p=0}^{m-1}\Pi\biggl\{ A_i^p(\b{r})B_i(\b{r})A_i^{m-p-1}(\b{r})
\biggr\}\, ,
\eq$$ where
$$
\eqalign{&A_i(\b{r})\equiv\!\!\!\!\!\sum_{\st j,k=1\atop\st
i\not=j\not=k\not=i}^N\!\!\!\frac{-1}{r_{jk}r_{ki}}K_{ijk}+\sum_{\st
j=1\atop
\st j\not=i}^N\sgn{i-j}K_{ij}
      +\sum_{\st j=1\atop\st j\not=i}^N\frac{1}{r_{ij}^2}+r_i^2 \cr
  &B_i(\b{r})\equiv\sum_{\st j=1\atop\st j\not=i}^N\biggl\{
\frac{1}{r_{ij}}\Bigl( \frac{\d}{\d r_i}+\frac{\d}{\d r_j}\Bigr)      -
\frac{1}{r_{ij}^2}\biggr\}K_{ij} \, .
}
\eq$$
The first two members of this set can be calculated  as
$$\eqalignno{
 H_2&=\sump{i,j=1}{N}\frac{1}{x_{ij}x_{ji}}P_ij \eqnumlabel\hdeux\cr
 H_4&=\sump{i,j,k\ell=1}{N}\biggl\{\frac{-1}{x_{ij}x_{jk}x_{k\ell}x_{\ell
i}}\biggr\}P_{ijk\ell}\cr
&\qquad\qquad\qquad+\sump{i,j=1}{N}\biggl\{\frac{2}{x_{ij}^4}-\frac{2}{3}
\frac{x_ix_j}{x_{ij}^2}-\frac{8}{3}(N-1)\frac{1}{x_{ij}^2}-\frac{4}{3}\biggr\}P
_{ij}\, . \eqnumlabel\hquatre
}$$
Because $H_2=2H^{\hbox{\scriptsize (FP)}}$, these
involutive quantities are necessarily invariants of the FP model. Note that
in the
calculation of  $H_4$, we have used the summation identities  (B.8), (B.15)
and (B.16) in
order to simplify the final result. We have also been able to crosscheck  the
commutativity of $H_2$ and
$H_4$
 by a direct computation, the details of which are given in Appendix A. It
should also
be pointed out that since it is impossible in quantum mechanics to freeze
the particle
positions onto the lattice sites and enforce at the same time the
vanishing of their momenta,
the absence of derivatives in $H_2$ and $H_4$ should not be regarded as a
consequence  of the
freezing procedure but as a rather impressive  mathematical cancelation,
whose {\it raison
d'\^{e}tre\/} has yet to be determined. Also, notice that our limiting
procedure does not
generate any odd-type conservation laws because the dynamical  Calogero
model simply does not
possess such symmetries. On the other hand, it is easy to see that the FP
model does possess
such symmetries by verifying explicitly that it commutes with the following operators
$$
\eqalignno{
 &H_1=\sum_{i=1}^N\frac{\d}{\d x_i}  \eqnumlabel\hun\cr
 &H_3=\sump{i,j,k=1}{N}\Bigl(
\frac{1}{x_{ij}x_{jk}x_{ki}}\Bigr)P_{ijk}-\frac{3}{2}\sump{i,j=1}{N}\Bigl(
\frac{1}{x_{ij}x_{ji}} \Bigr)\frac{\d}{\d x_i} \, .
   \eqnumlabel\htrois
}
$$
Here, we have not performed any simplifications on $H_3$, in order to show
that $\com{H_1}{H_3}=0$
(recall that the summation identities can only be used once all derivatives
have been commuted to the right).
It therefore seems possible to generate odd-type $H_m$, albeit by brute
force.  These odd conservation
laws seem to commute amongst themselves as well as with the even $H_m$
although they manifestly do not
possess the yangian symmetry. This means that Haldane and Talstra's
argument cannot be used to isolate their generating function; we have not
yet found the generating function for such a set.

\newsec{Conclusion}
Using an $\hbar$-expansion of the dynamical Calogero model, we have
succeeded in constructing
an even set $\{H_2, H_4 \dots\}$ of involutive charges for the
Frahm-Polychronakos spin chain, following to a large extent the procedure
of [\talhal]. However, as these authors pointed out, we should stress that
the underlying  $\hbar$-expansion constitutes a somewhat ad-hoc procedure
and does not seem to shed much light on the fundamental origin of these
conservation laws. One wonders whether the complicated limiting procedure
is really necessary and
whether these  invariants could not be generated in a simpler way, from an
intrinsic
spin-chain formulation.  In addition, we could ask whether explicit
expressions for these
Hamiltonian conservation laws could be written, in analogy with those of
the XXX model
[\ref{M. Grabowski and P. Mathieu, Ann. Phys. {\bf 243} (1995) 299; V.V.
Anshelevitch, Theo.
Mat. Phys. {\bf 43} (1980) 350.}].  We definitely see a similar pattern
emerging but the
expressions for the relative coefficients of the various terms appear
rather complicated.
Finally, a  brute force computation of $H_1$ and $H_3$ seems to hint at the
existence of an
odd set of involutive charges which does not obey the yangian symmetry, and
for which we still
lack a generating function.

\appendix{A}{ A direct computation of $\com{H_2}{H_4}$}
In this appendix, we show that the commutator $\com{H_2}{H_4}$  vanishes by
calculating
it explicitly. For compactness, let us start by expressing the conservation
laws in the form
$$\eqalignno{
  &H_2=\sump{i,j=1}{N}h_{ij}P_{ij} \eqnumlabel\hdeuxdeux\cr
 & H_4=-\sump{i,j,k,\ell=1}{N}h_{ijk\ell}P_{ijk\ell}+\sump{i,j=1}{N}f_{ij}P_{ij}\, ,
\eqnumlabel\hquatrequatre }$$
where
$$\eqalignno{
    & h_{ij}=\frac{1}{x_{ij}x_{ji}} \eqnumlabel\hij \cr
 &  h_{ijk\ell}=\frac{1}{x_{ij}x_{jk}x_{k\ell}x_{\ell i}}\eqnumlabel\hijk\cr
&
f_{ij}=\frac{2}{x_{ij}^4}-\frac{2}{3}\frac{x_ix_j}{x_{ij}^2}-\frac{8}{3}(N-1
)\frac{1}{x_{ij}^2}-\frac{4}{3}\, .
\eqnumlabel\fij
}$$
A  direct calculation  yields the following commutator
$$\eqalign{
\com{H_2}{H_4}=&\,\, 8\sump{i,j,k,\ell,m}{N} (h_{ij}-h_{im})
h_{jk\ell m}P_{ijk\ell m}\cr
&-4\sump{i,j,k,\ell}{N}(h_{ik}-h_{j\ell})h_{ijk\ell}P_{ij}P_{k\ell}\cr
&-4\sump{i,j,k}{N}
\biggl\{2\sum_{\ell=1\atop\sst\ell\not=i,j,k}^N(h_{i\ell}-h_{k\ell})
h_{ijk\ell}-(h_{ik}-h_{jk})f_{ij}\biggr\}P_{ijk}\,
. }\eqlabel\comuthdeuxquatre$$ Defining now the cyclic sum operator
$$\eqalign{
       \sum_{\{i_1\dots i_k\}}^{\hbox{\scriptsize cyclic}}f(x_{i_1}\dots
x_{i_k})
=f(x_{i_1}\dots x_{i_k})&+f(x_{i_2}, x_{i_3}\dots x_{i_k}, x_{i_1})\cr
&+\dots+f(x_{i_k}, x_{i_1}\dots x_{i_{k-1}})
}\eq$$
and using the fact that the exchange operators in \comuthdeuxquatre\ are
invariant under cyclic permutations of their
indices, we can write
$$\eqalign{
\com{H_2}{H_4}=&\,\, 8\sump{i,j,k,\ell,m}{N}\frac{1}{5}
\sum_{\{i,j,k,\ell,m\}}^{\hbox{\scriptsize cyclic}}\biggl\{
(h_{ij}-h_{im})h_{jk\ell m}\biggr\}P_{ijk\ell m}\cr
&-4\sump{i,j,k,\ell}{N}  \frac{1}{4}\sum_{\{i,j\}}^{\hbox{\scriptsize
cyclic}}\sum_{\{k,\ell\}}^{\hbox{\scriptsize
cyclic}}\biggl\{(h_{ik}-h_{j\ell})h_{ijk\ell}\biggr\}P_{ij}P_{k\ell}\cr
&-4\sump{i,j,k}{N}\frac{1}{3}\sum_{\{i,j,k\}}^{\hbox{\scriptsize
cyclic}}\biggl\{
2\sum_{\ell=1\atop\sst\ell\not=i,j,k}^N(h_{i\ell}-h_{k\ell})h_{ijk\ell}-(h_{
ik}-h_{jk})f_{ij}\biggr\}P_{ijk}\, .}\eqlabel\commdeux$$ 
This commutator will therefore vanish if one can prove that
$$\eqalignno{
 &F_1\equiv\sum_{\{i,j,k,\ell,m\}}^{\hbox{\scriptsize cyclic}}\biggl\{
(h_{ij}-h_{im})h_{jk\ell m}\biggr\}=0\eqnumlabel\condcomun\cr
&F_2\equiv\sum_{\{i,j\}}^{\hbox{\scriptsize
cyclic}}\sum_{\{k,\ell\}}^{\hbox{\scriptsize
cyclic}}\biggl\{(h_{ik}-h_{j\ell})h_{ijk\ell}\biggr\}=0\eqnumlabel
\condcomdeux\cr
&F_3\equiv\sum_{\{i,j,k\}}^{\hbox{\scriptsize
cyclic}}\biggl\{-2\sum_{\ell=1\atop\sst\ell\not=i,j,k}^N(h_{i\ell}-h_{k\ell}
)h_{ijk\ell}+
(h_{ik}-h_{jk})f_{ij}\biggr\}=0\, .\eqnumlabel\condcomtrois
}$$ The first condition \condcomun\ is shown to be satisfied in the
following manner. First,
we extract
 a  cyclic invariant from the sum
$$\eqalign{
  F_1&=\sum_{\{i,j,k,\ell,m\}}^{\hbox{\scriptsize
cyclic}}\biggl\{\biggl(\frac{1}{x_{ij}^2}-\frac{1}{x_{im}^2}\biggr)\frac{1}{
x_{jk}x_{k\ell}x_{\ell m}x_{mj}}\biggr\}\cr
&=\frac{1}{x_{ij}x_{jk}x_{k\ell}x_{\ell
m}x_{mi}}\sum_{\{i,j,k,\ell,m\}}^{\hbox{\scriptsize
cyclic}}\biggl\{\frac{x_{mi}}{x_{ij}x_{mj}}-\frac{x_{ij}}{x_{mi}x_{mj}}
\biggr\}\cr
&\equiv h_{ijk\ell m}\sum_{\{i,j,k,\ell,m\}}^{\hbox{\scriptsize
cyclic}}\biggl\{\frac{x_{mi}}{x_{ij}x_{mj}}-\frac{x_{ij}}{x_{mi}x_{mj}}\biggr\}\, .
 \cr
}\eq$$
The next few steps are just basic  algebra
$$\eqalign{
F_1&=h_{ijk\ell m}\sum_{\{i,j,k,\ell,m\}}^{\hbox{\scriptsize
cyclic}}\biggl\{\frac{x_{mi}^2-x_{ij}^2}{x_{ij}x_{mj}x_{mi}}\biggr\}\cr
&=h_{ijk\ell m}\sum_{\{i,j,k,\ell,m\}}^{\hbox{\scriptsize
cyclic}}\biggl\{\frac{(x_m+x_j)x_{mj}-2x_ix_{mj}}{x_{ij}x_{mj}x_{mi}}\biggr\}\cr
&=h_{ijk\ell m}\sum_{\{i,j,k,\ell,m\}}^{\hbox{\scriptsize
cyclic}}\biggl\{\frac{x_m+x_j-2x_i}{x_{ij}x_{mi}}\biggr\}\, .\cr }\eq$$
Extracting once more the
cyclic invariant $h_{ijk\ell m}$ gives
$$\eqalign{
&F_1=h_{ijk\ell m}^2\sum_{\{i,j,k,\ell,m\}}^{\hbox{\scriptsize
cyclic}}\biggl\{(x_m+x_j-2x_i)x_{jk}x_{k\ell}x_{\ell m}\biggr\}\, .\cr }\eq$$
This cyclic sum can then be shown to vanish by plainly writing down all of
its terms.
The second condition  \condcomdeux\ can be proved to hold in a similar fashion.
Establishing
the vanishing of $F_3$ is a bit more tricky however. The main steps are as
follows. First, we
write $F_3$ explicitly:
$$\eqalign{
 F_3=\sum_{\{i,j,k\}}^{\hbox{\scriptsize cyclic}}\biggl\{
\frac{2}{x_{ij}x_{jk}}&\biggl[\sum_{\ell=1\atop\sst \ell\not=i,k}\biggl(
\frac{1}{x_{k\ell}x_{\ell i}^3}-\frac{1}{x_{i\ell}x_{\ell
k}^3}\biggr)-\frac{1}{x_{kj}x_{ji}^3}+\frac{1}{x_{ij}x_{jk}^3}\biggr]\cr
&+\biggl(
\frac{1}{x_{jk}^2}-\frac{1}{x_{ik}^2}\biggr)\biggl(\frac{2}{x_{ij}^4}-\frac{
2}{3}\frac{x_ix_j}{x_{ij}^2}-\frac{8}{3}(N-1)\frac{1}{x_{ij}^2}-\frac{4}{3}
\biggr)\biggr\}\, .\cr
}\eq$$
We start by using the summation identity (B.17) in
order to simplify the first two terms and  notice that the  last two terms
in the last
parenthesis do not contribute. The reduced expression is:
$$\eqalign{
F_3=\sum_{\{i,j,k\}}^{\hbox{\scriptsize cyclic}}\biggl\{
\frac{2}{x_{ij}x_{jk}}\biggl[\frac{1}{3}&\frac{x_k^2-x_i^2}{x_{ki}^2}-\frac{
1}{2}\frac{x_i+x_k}{x_{ki}}-\frac{1}{x_{kj}x_{ji}^3}+\frac{1}{x_{ij}x_{jk}^3
}\biggr]\cr
&+\frac{2}{x_{ij}^4x_{jk}^2}-\frac{2}{x_{ij}^4x_{ik}^2}+\frac{2}{3}\frac{x_i
x_j}{x_{ij}^2x_{ik}^2}-\frac{2}{3}\frac{x_ix_j}{x_{ij}^2x_{jk}^2}\biggr\}
\, .\cr
}\eq$$
The cyclic sum will also cancel the last two  terms in square brackets with
the subsequent two
terms so that we are left with
$$\eqalign{
F_3&=\sum_{\{i,j,k\}}^{\hbox{\scriptsize
cyclic}}\biggl\{ -\frac{1}{3}\frac{x_i+x_k}{x_{ij}x_{jk}x_{ki}}
+\frac{2}{3}\frac{x_ix_j}{x_{ij}^2x_{ik}^2}-\frac{2}{3}\frac{x_ix_j}{x_{ij}^
2x_{jk}^2}\biggr\}\cr
&=-\frac{1}{3}\frac{1}{x_{ij}x_{jk}x_{ki}}\sum_{\{i,j,k\}}^{\hbox{\scriptsize
cyclic}}(x_i+x_k)+\frac{2}{3}\frac{1}{x_{ij}^2x_{jk}^2x_{ki}^2}\sum_{\{i,j,k
\}}^{\hbox{\scriptsize
cyclic}}x_ix_j(x_{jk}^2-x_{ki}^2)\cr
&=-\frac{2}{3}\frac{(x_i+x_j+x_k)}{x_{ij}x_{jk}x_{ki}}+\frac{2}{3}\frac{(x_i
x_j^3-x_i^3x_j+x_jx_k^3-x_j^3x_k+x_kx_i^3-x_ix_k^3)}{x_{ij}^2x_{jk}^2x_{ki}^
2}\cr
&=-\frac{2}{3}\frac{(x_i+x_j+x_k)}{x_{ij}x_{jk}x_{ki}}+
\frac{2}{3}\frac{(x_i+x_j+x_k)x_{ij}x_{jk}x_{ki}}{x_{ij}^2x_{jk}^2x_{ki}^2}=0\, .
}\eq$$
We therefore see that $H_2$ and $H_4$ do commute, a fact which corroborates
the validity of the dynamical $\hbar$-expansion used throughout this work.

\appendix{B}{The zeroes of the Hermite polynomials: summation identities}

In this appendix, we briefly show how the lattice sites of the FP model,
defined by \sumpube, can be identified with the zeroes of
the Hermite polynomial $H_N(x)$ and then present a series of summation
identities which are vital for the reduction of certain
expressions. Following [\frahm], consider then the Hermite differential equation
$$
  H''_N(x)-2xH'_N(x)+2NH_N(x)=0\, .
\eqlabel\hermdiff$$
Letting $x_i (i=1\dots N)$ denote the zeroes of $H_N(x)$ and  evaluating
\hermdiff\  at an arbitrary zero $x_{\ell}$ gives
$$
    H_N''(x_{\ell})=2x_{\ell} H'_N(x_{\ell})  \, .
\eqlabel\hermcond$$
Factorizing $H_N(x)$ in terms of its zeroes and substituting in \hermcond\
generates the identity
$$
   \sum_{\st k=1\atop\st k\not=j}^N\frac{1}{x_{jk}}=x_j \, .
\eqlabel\rela$$
As already mentioned, a number of  simple summation identities
generalizing the previous one  have already been discovered some time ago
[\zeroesi]. One can easily generate more complicated formulae. The general
procedure is the following:  to increment a power to the numerator of \rela
, we can
proceed as follows
$$\eqalign{
   \sum_{\st k=1\atop\st k\not=j}^N\frac{x_{jk}}{x_{jk}}=(N-1)\quad
&\Longrightarrow\quad x_j \sum_{\st k=1\atop\st k\not=j}^N\frac{1}{x_{jk}}-
\sum_{\st k=1\atop\st k\not=j}^N\frac{x_k}{x_{jk}}=(N-1)\cr
&\Longrightarrow\quad
 \sum_{\st k=1\atop\st k\not=j}^N\frac{x_k}{x_{jk}}=x_j^2-(N-1)\, .
}\eqlabel\rela$$
On the other hand, to increase a power in the denominator, the procedure is
$$
\eqalign{
&\sum_{\st k=1\atop\st k\not=j}^N\frac{1}{x_{jk}}-\sum_{\st k=1\atop\st
k\not=i}^N\frac{1}{x_{ik}} = \sum_{\st k=1\atop\st
k\not=i,j}^N\frac{x_{ik}-x_{jk}}{x_{jk}x_{ik}} + \frac{1}{x_{ji}}-
\frac{1}{x_{ij}}=
\sum_{\st k=1\atop\st
k\not=i,j}^N\frac{x_{ij}}{x_{jk}x_{ki}}- \frac{2}{x_{ij}}\Longrightarrow\cr
& \sum_{\st k=1\atop\st
k\not=i,j}^N\frac{1}{x_{jk}x_{ik}}= -\frac{1}{x_{ij}}\left[\sum_{\st k=1\atop\st
k\not=j}^N\frac{1}{x_{jk}}- \sum_{\st k=1\atop\st k\not=i}^N\frac{1}{x_{ik}}
+\frac{2}{x_{ij}}\right] =  -\frac{1}{x_{ij}}\left[
x_j-x_i+\frac{2}{x_{ij}}\right]\, ,\cr}\eq$$
that is
$$ \sum_{\st k=1\atop\st
k\not=i,j}^N\frac{1}{x_{jk}x_{ik}}= 1-\frac{2}{x^2_{ij}}\, .\eq$$

Using such procedures, one can generate a whole set of summation
identities, the most useful of them being given by
$$\eqalignno{
&  \sum_{\st k=1\atop\st k\not=i}^N\frac{1}{x_{ik}}=x_i \eqnumlabel\relb\cr
&   \sum_{\st k=1\atop\st
k\not=i}^N\frac{1}{x_{ik}^2}=\frac{2}{3}(N-1)-\frac{1}{3}x_i^2
\eqnumlabel\relc\cr
&     \sum_{\st k=1\atop\st k\not=i}^N\frac{1}{x_{ik}^3}=\frac{1}{2}x_i
\eqnumlabel\reld\cr
&     \sum_{\st k=1\atop\st k\not=i}^N\frac{1}{x_{ik}^4}=\frac{1}{45}\Bigl[
2(N+2)-x_i^2 \Bigr] \Bigl[ 2(N-1)-x_i^2 \Bigr] \eqnumlabel\rele\cr
&    \sum_{\st k=1\atop\st k\not=i}^N\frac{x_k}{x_{ik}}=x_i^2-(N-1)
\eqnumlabel\relf\cr
&    \sum_{\st k=1\atop\st
k\not=i}^N\frac{x_k}{x_{ik}^2}=-\frac{1}{3}x_i^3+\Bigl[\frac{2}{3}(N-1)-1
\Bigr] \eqnumlabel\relg\cr
&  \sum_{\st k=1\atop\st
k\not=i}^N\frac{x_k}{x_{ik}^3}=\frac{5}{6}x_i^2-\frac{2}{3}(N-1)
\eqnumlabel\relh\cr
&  \sum_{\st k=1\atop\st k\not=i}^N\frac{x_k^2}{x_{ik}}=x_i^3-(N-2)x_i
\eqnumlabel\reli\cr
&  \!\!\!\!\!\sum_{\st k=1\atop\st
i\not=j\not=k\not=i}^N\!\!\!\frac{1}{x_{ik}x_{kj}}=1-\frac{2}{x_{ij}^2}
\eqnumlabel\relj\cr
&  \!\!\!\!\!\sum_{\st k=1\atop\st
i\not=j\not=k\not=i}^N\!\!\!\frac{1}{x_{ik}x_{kj}^2}=\frac{1}{3}\frac{2N+1-x
_j^2}{x_{ij}}-\frac{3}{x_{ij}^3}\eqnumlabel\relk\cr
&  \!\!\!\!\!\sum_{\st k=1\atop\st
i\not=j\not=k\not=i}^N\!\!\!\frac{1}{x_{ik}x_{kj}^3}=\frac{1}{3}\frac{2N+1-x
_j^2}{x_{ij}^2}-\frac{1}{2}\frac{x_j}{x_{ij}}- \frac{4}{x_{ij}^4}
\eqnumlabel\relm\cr
}$$
We now notice that \reld\ is identical to \sumpube ,  the condition for the
FP model to be $Y[su(n)]$-symmetric. This proves
that the lattice sites of the FP model with $N$ spins are actually the
zeroes of the $H_N(x)$ polynomial.

\vskip0.3cm
\centerline{\bf Acknowledgment}

We thank F. Lesage and L. Lapointe for useful discussions and CRM for its
hospitality during  the period in which this work was completed.

\vskip 0.5cm
\centerline{\bf REFERENCES}

\immediate\closeout\refs \vskip 0.5cm
  \message{References}\input references
\vfill\eject

\end